\documentclass[preprint2]{aastex}
\usepackage{psfig}
\include{epsf}
\newcommand{\be}{\begin{equation}}
\newcommand{\ee}{\end{equation}}
\newcommand{\ba}{\begin{eqnarray}}
\newcommand{\ea}{\end{eqnarray}}

\def\simless{\mathbin{\lower 3pt\hbox
   {$\rlap{\raise 5pt\hbox{$\char'074$}}\mathchar"7218$}}}
\def\simgreat{\mathbin{\lower 3pt\hbox
   {$\rlap{\raise 5pt\hbox{$\char'076$}}\mathchar"7218$}}}   

\shorttitle{SCUBA/Spitzer Investigation of Submm/far-IR Background}
\shortauthors{S. Dye et al.}

\begin{document}


\title{A SCUBA/Spitzer Investigation of the Far Infra-Red Extragalactic 
Background}

\author{S. Dye\altaffilmark{1}, S. A. Eales\altaffilmark{1},
M. L. N. Ashby\altaffilmark{2}, J. -S. Huang\altaffilmark{2},
E. Egami\altaffilmark{3},
M. Brodwin\altaffilmark{4},
S. Lilly\altaffilmark{5}, 
T. M. A. Webb\altaffilmark{6}}
\affil{}
\altaffiltext{1}{School of Physics \& Astronomy, Cardiff University, 
5 The Parade, Cardiff, CF24 3YB, U.K.}
\altaffiltext{2}{Harvard Smithsonian Centre for Astrophysics, 60 Garden Street,
Cambridge, MA 02138, U.S.A.}
\altaffiltext{3}{Steward Observatory, University of Arizona, 933
North Cherry Avenue, Tuscon, AZ 85721, U.S.A.}
\altaffiltext{4}{JPL, CalTech, M/S 169-506, 4800 Oak Grove Drive, 
Pasadena, CA 91109, U.S.A.}
\altaffiltext{5}{Institute of Astronomy, Swiss Federal Institute of 
Technology Zurich, CH-8093 Zurich, Switzerland}
\altaffiltext{6}{Sterrewacht Leiden, Neilss Bohrweg 2, Leiden 233CA, 
Netherlands}

\label{firstpage}

\begin{abstract}
We have measured the contribution of submillimeter and mid-infrared
sources to the extragalactic background radiation at 70 and
160$\mu$m. Specifically, we have stacked flux in 70 and 160$\mu$m
Spitzer Space Telescope ({\sl Spitzer}) observations of the Canada-UK
Deep Sub-millimeter Survey 14h field at the positions of 850$\mu$m
sources detected by SCUBA and also 8 and 24$\mu$m sources detected by
{\sl Spitzer}.  We find that per source, the SCUBA galaxies are the
strongest and the 8$\mu$m sources the weakest contributors to the
background flux at both 70 and 160$\mu$m. Our estimate of the
contribution of the SCUBA sources is higher than previous
estimates. However, expressed as a total contribution, the
full 8$\mu$m source catalogue accounts for twice the total 24$\mu$m
source contribution and $\sim 10$ times the total SCUBA source
contribution.  The 8$\mu$m sources account for the majority of the
background radiation at 160$\mu$m with a flux of 0.87$\pm$0.16 MJy/sr
and at least a third at 70$\mu$m with a flux of 0.103$\pm$0.019
MJy/sr. These measurements are consistent with current lower limits on
the background at 70 and 160$\mu$m.  Finally, we have investigated the
70 and 160$\mu$m emission from the 8 and 24$\mu$m sources as a
function of redshift. We find that the average 70$\mu$m flux per
24$\mu$m source and the average 160$\mu$m flux per 8 and 24$\mu$m
source is constant over all redshifts, up to $z\sim 4$. In contrast,
the low-redshift half $(z<1)$ of the of 8$\mu$m sample contributes
approximately four times the total 70$\mu$m flux of the high-redshift
half. These trends can be explained by a single non-evolving SED.

\end{abstract}


\section{Introduction}

Excluding the microwave background, approximately half of the entire
extragalactic background radiation is emitted by dust at far infra-red
(IR) and sub-millimeter (submm) wavelengths
\citep[e.g.,][]{fixsen98,hauser01,dole06}.  This cosmic IR background
(CIB) radiation peaks at a wavelength of $\sim 200\mu$m, yet compared
to the optical, relatively little is known about the sources
responsible.

Surveys conducted by the Sub-millimeter Common User Bolometer Array
(SCUBA) and the Max-Planck Millimeter Bolometer (MAMBO) over the last
decade have directly resolved up to two-thirds of the CIB at $850\mu$m
and 1.1mm into discrete, high redshift sources (although this fraction
is uncertain due to the uncertainty in measurements of the CIB at
these wavelengths). Discovery of this population has been extremely
important since SCUBA galaxies represent the most energetic
star-forming systems at an epoch when the Universe was at its most
active. However, the impact this has had on understanding the nature
of the CIB is relatively minor since at these wavelengths, the CIB has
30 to 40 times less power than at the peak.

A recent study using a large sample of 73 bright ($\simgreat 5$mJy)
SCUBA sources by \citet{chapman05} indicated that the population
contributes a mere $\sim 2\%$ of the CIB at the peak, with an
extrapolation of up to $\sim 6\%$ for sources down to the fainter
limit of 1mJy.  However, this work relied on an assumed spectral
energy distribution (SED) for the SCUBA sources constrained only by a
redshift, the 850$\mu$m SCUBA flux and a radio flux at
1.4GHz. Furthermore, redshifts were obtained from optical spectra
having identified the optical sources with radio counterparts to the
SCUBA sources.  This introduces two selection effects. The first
causes SCUBA sources with $z \simgreat 3$ to be missed by requiring a
radio detection, the selection function for radio sources falling off
rapidly at $z \sim 3$ due to the K-correction.  The second causes a
paucity of sources around $z \sim 1.5$ where no emission lines fall
within the observable wavelength range of their spectra.

This motivates the first of two main goals of this paper. By stacking
the flux in 70 and 160$\mu$m MIPS images at the positions of SCUBA
sources detected in the Canada-United Kingdom Deep Sub-millimeter
Survey (CUDSS) 14-hour field \citep{eales00,webb03}, we directly
measure the contribution of the SCUBA sources to the CIB at
wavelengths in the vicinity of the peak.

In addition to the submm surveys, space-borne mid-IR surveys conducted
by the Infra-red Astronomical Satellite (IRAS) and the Infra-red Space
Observatory (ISO) have resolved significant contributions to the CIB
from the shorter wavelength side of the peak \citep[see, for example,
the review by][]{lagache05}. The introduction of the Spitzer Space
Telescope \citep[{\sl Spitzer};][]{werner04} means that such surveys
can be carried out over much wider areas and to much greater depths.

In particular, the Multi-band Photometer for {\sl Spitzer}
\citep[MIPS;][]{rieke04} has been used for a variety of mid- and
far-IR surveys to resolve sources contributing to the CIB.
\citet{papovich04} showed that approximately 70\% of the CIB at
24$\mu$m can be resolved into IR galaxies with flux $\ge 60\mu$Jy. In
contrast, \citet{dole04} found that at the longest two MIPS
wavelengths, 70 and 160$\mu$m, only 20\% and 10\% of the CIB can be
directly resolved into distinguishable sources brighter than 3.2 and
40mJy respectively. However, the error on these fractional quantities is
very large since the absolute flux of the background at 160$\mu$m
is currently unknown to a factor of $\sim 2$ and at 70$\mu$m, the
uncertainty is even larger.

A major problem with attempting to directly resolve sources in deep 70
and 160$\mu$m MIPS surveys is source confusion due to the large
instrument point spread function (PSF). This problem can be
circumvented by measuring the 70 and 160$\mu$m MIPS flux at the
position of objects selected in other wavebands for which there are
already accurate positions. This stacking technique has been
successfully used by several authors with SCUBA data that also suffer
from confusion
\citep[e.g.,][]{peacock00,serjeant04,knudsen05,dye06,wang06}.
\citet{dole06} stack MIPS flux at the positions of 24$\mu$m sources
with fluxes $> 60\mu$Jy to find that they represent the bulk of
the CIB at 70 and 160$\mu$m respectively (see Section
\ref{sec_full_stack_results}). These contributions are investigated as
a function of 24$\mu$m source flux and, based on external studies of
the redshift distribution of MIPS 24$\mu$m sources, the authors
conclude that the majority of the radiation must be emitted at $z \sim
1$.

This provides the second main motivation for the present paper.  We
extend the analysis of \citet{dole06} in two ways. Firstly, we
additionally measure the contribution to the CIB at 70 and 160$\mu$m
from 8$\mu$m sources observed with {\sl Spitzer}'s Infra-red Array Camera
\citep[IRAC;][]{fazio04}. Secondly, we investigate how the
contribution from the 8 and 24$\mu$m populations varies with redshift,
using photometric redshifts established for these sources in our
earlier work \citep{dye06}. 

This paper is set out as follows. In the following section we describe
the data. Section \ref{sec_analysis} outlines our stacking procedure.
Our results are presented in Section \ref{sec_results}, followed
by a summary and brief discussion in Section \ref{sec_summary}.

\section{Data}

Coverage of the CUDSS 14h field in this paper comprises 850$\mu$m
SCUBA observations as well as data acquired with both {\sl Spitzer}'s
IRAC and MIPS instruments.  The photometric redshifts of the 24 and
8$\mu$m sources used later were determined from ground-based U, B, V,
I and K photometry as well as IRAC 3.6$\mu$m and 4.5$\mu$m
observations. We refer the reader to \citet{dye06} for a full account
of the determination of these redshifts.

\subsection{SCUBA data}

The SCUBA catalogue contains sources extracted from 63 hours worth of
850$\mu$m data taken on 20 different nights over the period from March
1998 to May 1999 at the James Clark Maxwell Telescope (JCMT). The
850$\mu$m map of the $\sim 7' \times 6'$ survey region was composed by
combining several jiggle maps at different base positions. Each jiggle
map was observed for approximately one hour with a 64-point 
pattern (to ensure full sampling), nodding JCMT's secondary mirror and
chopping by $30''$ in right ascension. We refer the reader to
\citet{eales00} for more specific details of the data reduction.

The source list used for the stacking is that compiled from the
850$\mu$m data by \citet{webb03}, consisting of 23 sources above a $3
\sigma$ detection threshold within the 41 arcmin$^2$ SCUBA map. The
average $3 \sigma$ sensitivity limit of the sample is 3.5mJy. 20 of
these sources lie within the MIPS 70$\mu$m coverage and 22 within the
160$\mu$m coverage.

\subsection{Spitzer Space Telescope Data}
\label{sec_spit_data}

\begin{figure}
\epsfxsize=8.4cm
{\hfill
\epsfbox{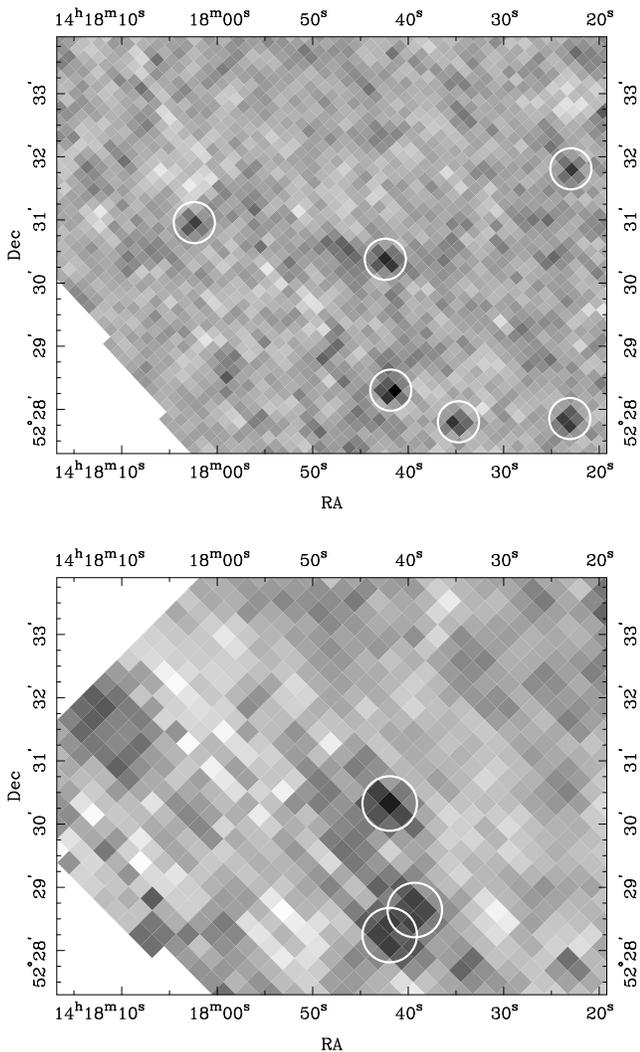}
\hfill}
\epsfverbosetrue
\caption{\small MIPS observations of the CUDSS 14h field at 70$\mu$m
(top) and 160$\mu$m (bottom). Sources detected with $\geq 3\sigma$
significance are circled. Blank regions are masked areas of
poor or no coverage.}
\label{mips_maps}
\end{figure}

The {\sl Spitzer} observations discussed in this paper were obtained as part
of the Guaranteed Time Observing program number 8 to image the
extended Groth strip, a $2^{\circ} \times 10'$ area at $\alpha \sim
14^{\rm h} 19^{\rm m}$, $\delta \sim 52^{\circ}48'$ (J2000) with IRAC
and MIPS. In the present work, we have limited the stacking to a small
section of the extended Groth strip that fully contains the CUDSS 14
hour field. This ensures a self-consistent comparison between the
SCUBA, 24$\mu$m and 8$\mu$m  source stacking.  Of this section, 96\% of
the CUDSS 14 hour field falls inside the 70$\mu$m coverage and 91\%
inside the 160$\mu$m coverage. The south-east corner of the 70$\mu$m
data and the south-east and north-east corners of the 160$\mu$m data
have either poor or no coverage and these are masked out in all
analyses throughout this paper. Figure \ref{mips_maps} shows the
images.

The 24 and 8$\mu$m source catalogues used for stacking were first
presented in \citet{ashby06} and we refer the reader to this work for
a detailed account of their creation. Both catalogues cover a slightly
larger area than the original CUDSS 14 hour field but are contained by
the 70 and 160$\mu$m image sections described above. The 24$\mu$m
sources cover an area of 49 arcmin$^2$ and all lie above a $5\sigma$
point source sensitivity of 70$\mu$Jy.  The 8$\mu$m sources cover 59
arcmin$^2$ and lie above a $5\sigma$ point source sensitivity of $5.8
\mu$Jy. There are a total of 177 24$\mu$m sources that lie within the
MIPS coverage at 70$\mu$m and 171 within the 160$\mu$m coverage.  Of
the 8$\mu$m sources, 801 and 773 lie within the MIPS coverage at
70$\mu$m and 160$\mu$m respectively.

The MIPS 70 and 160$\mu$m images were observed in scan map mode with
the slow scan rate. The data were processed with the Spitzer
Science Centre (SSC) pipeline
\citep{gordon05} to produce images with flux measured in MIPS
instrumental units. These were converted to units of mJy/arcsec$^2$
using the calibration factors 14.9 mJy/arcsec$^2$ per data unit for
the 70$\mu$m data and 1.0 mJy/arcsec$^2$ per data unit for the
160$\mu$m data. Note that these are $5-10\%$ smaller than those quoted
for the MIPS-Ge pipeline in the MIPS data handbook (version 3.2.1, 6
Feb 2006 release\footnote{see {\em
http://ssc.spitzer.caltech.edu/mips/dh}}) since the SSC pipeline is
completely independent.  The pixel size in the 70 and 160$\mu$m images
is respectively $9.85'' \times 9.85''$ and $16.00'' \times 16.00''$.
For comparison, the FWHM (full width at half max) of the instrumental
PSFs are $\sim 16''$ at 70$\mu$m and $\sim 40''$ at 160$\mu$m
(measured by fitting to the central Gaussian component of the
PSF). The data have a $5\sigma$ point source sensitivity of 10mJy at
70$\mu$m and 60mJy at 160$\mu$m.

The error images output by the current version of the SSC pipeline
are only an estimate of the true error and do not accommodate for the
full range of effects exhibited by the MIPS detectors.  Also, the MIPS
image data are covariant as a result of pixel interpolation and
rebinning carried out during pipeline construction of the mosaics.
This covariance must be quantified and incorporated into the stacking
analysis that follows.  For these reasons, we generated our own error
data.

To obtain variance images, we made the assumption that the error in
the flux of a given pixel is inversely proportional to the square root
of the number of times the pixel has been scanned. Of course, pixels
containing bright sources will have an additional contribution from
Poisson noise, but since our data have very few bright sources and
since we investigate their effect on the stacking by removing them,
this is not a concern.  The variance images are then the inverse of
the coverage maps scaled to have a standard deviation of
unity. Indeed, we recover a perfect Gaussian distribution of pixel
signal to noise (apart from a few outliers due to bright sources) when
the variance is calculated in this manner.

Pixel covariances were derived directly from the image data.  We
calculated the average covariance for all pixel pair configurations
(up to a separation such that the covariance was negligible) over
image areas away from bright sources. Avoiding areas with bright
sources helps minimise the overestimation caused by the instrumental
PSF. Nevertheless, as we discuss in Section \ref{sec_analysis}, the
covariance is still overestimated by a small amount, meaning that our
quoted significances are conservative.

\section{Analysis}
\label{sec_analysis}

Rather than follow the procedure of stacking small sections of the
image centred on the source positions \citep[see, for
e.g., ][]{dole06}, we opt for the method used in our earlier work
\citep{dye06} whereby flux is measured directly from the image at each
source position. The catalogue of sources is offset by varying amounts
on a 2D regular grid and the flux summed over all sources at each
offset. The result is an `offset map' that gives an indication of
how well aligned the sources are with respect to the image and the
significance of the stacked flux (see Figure \ref{corr_maps}). As we
showed in \citet{dye06}, if the sources are properly aligned with
the image, then the correct stacked flux is that at the origin of the
offset map, not necessarily at the peak which may be slightly
offset from the origin.

The data stacked in \citet{dye06} were SCUBA maps with each pixel
value representing the total flux a point source would have if located
within that pixel. The stacking therefore simply took the sum of all
map pixel values at the source positions. In the current work, the
MIPS data output by the pipeline adhere to the usual optical
convention whereby a pixel holds the flux received solely by that
pixel.  A source's total flux is therefore the sum of flux in all
pixels belonging to the source. To convert the MIPS data into the
convention used by the SCUBA data in preparation for stacking, we
convolved the images with a circular top-hat, then multiplied them by
the aperture correction corresponding to the top-hat radius.  The
convolution was carried out at the original pixel scale of each image,
so to prevent aliasing effects, pixels around the top-hat circumference
were weighted by their interior fractional area.

Our choice of a circular top-hat instead of the more conventional
instrument PSF was based on the fact that the MIPS PSF varies between
sources and between images. We created simulated MIPS images of a
point source, varying the asymmetry and size of the image PSF compared
to the fiducial model PSF in each case. We found that the error in the
total source flux measured by convolving with the fiducial PSF rises
more quickly with increasing PSF asymmetry and size than measured by
convolving with a circular top-hat having an aperture correction
matched to the fiducial PSF. The MIPS data handbook recommends that
the PSF should be determined directly from bright sources in the
image, but since our image has no sufficiently bright sources, this
was not possible.

With this in mind, we chose top-hat radii of $r=18''$ and $r=40''$ for
the 70 and 160$\mu$m images respectively. Instead of using the
corresponding aperture corrections from the MIPS data handbook, we
computed our own to ensure consistency with our top-hat convolution.
For each wavelength, we took the in-orbit PSF\footnote{provided at
{\em http://ssc.spitzer.caltech.edu/mips/psf.html}}, binned it to the
relevant image pixel scale, then computed its product with the
edge-weighted top-hat to give the fraction of flux contained within
the top-hat and hence the aperture correction.  For the 70$\mu$m data,
we measured an aperture correction of 1.63 for the $r=18''$ top-hat
and for the 160$\mu$m data with the $r=40''$ top-hat, an aperture
correction of 1.53. These are $\sim 5\%$ smaller than the low
temperature aperture corrections given in the MIPS handbook.

In this paper, we quote an average stacked flux per source,
$\overline{f}=\Sigma_{i=1}^{N} f_i/N$ and an average inverse variance
weighted flux per source, $\overline{f}_w=\Sigma_i (f_i
\sigma_i^{-2})/ \Sigma_i \sigma_i^{-2}$. Here, $f_i$ and $\sigma_i$
are respectively the flux and $1\sigma$ uncertainty on the top-hat
convolved image pixel populated by source $i$. Comparison of the average
flux with the weighted average flux gives an indication of whether the
stacked signal is dominated by a minority of high significance
sources. Simple error propagation shows that $\sigma_i$ is 
given by
\be
\sigma_i^2 = \sum_{j,k} t_{j-i} t_{k-i} c_{jk}
\ee
where $t_i$ is the value of the top-hat function in pixel $i$
and $c_{jk}$ is the covariance in the original, unconvolved
image between pixels $j$ and $k$. 
As explained in Section \ref{sec_spit_data}, the variances,
i.e., diagonal terms of $c_{jk}$, come from the variance image,
computed for each image pixel from the coverage map. However, the
off-diagonal terms are the covariances averaged over the whole image, so
that any pixel pair $jk$ with the same separation vector are assigned
the same covariance.

To verify our treatment of errors, for each of the 70 and 160$\mu$m
data, we fitted a Gaussian to the distribution of flux significance in
the original unconvolved images, and to the distribution of the
significance, $f_i/\sigma_i$, for the convolved images.  In the latter
case, we first included, then omitted the off-diagonal elements in the
covariance matrix. With the original 70 and 160$\mu$m images, the
Gaussian fit had unit standard deviation as expected.  With the
convolved images and the off-diagonal covariance terms included, the
standard deviation for the 70$\mu$m image was approximately 0.95 and for
the 160$\mu$m image, 0.90.  However, including only the variance terms
gave a standard deviation of 1.79 for the 70$\mu$m data and 2.78 for
the 160$\mu$m data. This test confirms two facts: 1) If covariance is
not allowed for, the stacked flux error is underestimated by $\sim
45\%$ at 70$\mu$m and $\sim 65\%$ at 160$\mu$m. 2) Our measurement of
covariance is slightly overestimated, presumably due to the MIPS PSF,
giving rise to a conservative 5-10\% underestimate of the stacked flux
significance.


Finally, source confusion due to the large 70 and 160$\mu$m MIPS PSF
must also be accounted for in the stacking. With the nodded and
chopped SCUBA data of \citet{dye06}, this could be neglected because
the beam and hence the map in these data had an average of zero. With
the MIPS data, this is not so. The average stacked flux per source
must therefore be corrected by subtracting off the average of the
convolved image then dividing the result by the factor $(1-B_e/A)$,
where $B_e$ is the effective area of the PSF of the convolved image
and $A$ is the image area (see Appendix \ref{app_flux_deboost}).

\section{Results}
\label{sec_results}

To investigate the contribution from bright, directly detectable sources
in the MIPS images to the average stacked flux, we carried out two
stacks per image, one leaving the image unaltered and a second
with all $\geq 3 \sigma$ sources removed.  At 70$\mu$m, there are six
$\geq 3 \sigma$ sources, whereas at 160$\mu$m, there are three (see Figure
\ref{mips_maps}). Sources were removed by subtracting the in-orbit PSFs 
from the images at the position of each source, scaled to match the
integrated source brightness.

\subsection{Stacking the full SCUBA, 24$\mu$m and 8$\mu$m catalogues}
\label{sec_full_stack_results}

\begin{figure*}
\epsfxsize=15.8cm
{\hfill
\epsfbox{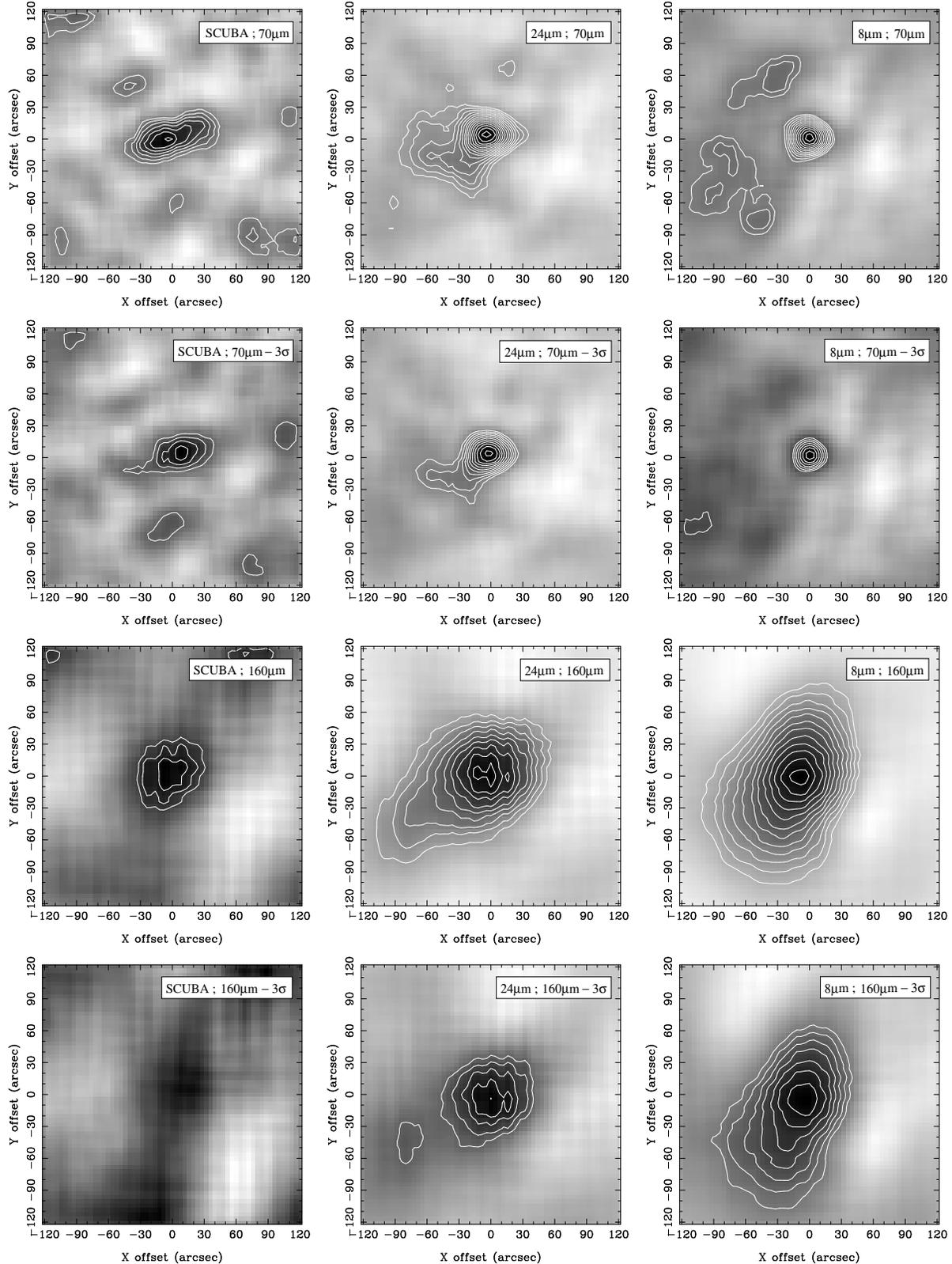}
\hfill}
\epsfverbosetrue
\caption{\small Offset maps of average weighted stacked flux per
source ($\overline{f}_w$) for all combinations
of MIPS image and sources stacked. Columns from left to right correspond
to SCUBA sources, 24$\mu$m sources and 8$\mu$m sources. First and
second rows correspond to 70$\mu$m MIPS data including then excluding
$\geq 3\sigma$ sources respectively. Similarly, third and fourth rows
correspond to 160$\mu$m MIPS data including then excluding the $\geq3
\sigma$ sources. Contours start at $2\sigma$ significance and increase
by $0.5\sigma$ intervals. Average fluxes for each case are given in
Table \ref{tab_fluxes}. Significances account for the MIPS 70 and
160$\mu$m calibration uncertainty. The number of sources stacked for each
of the different combinations are: 20 (SCUBA/70), 22 (SCUBA/160),
177 (24/70), 171 (24/160), 801 (8/70) and 773 (8/160).}
\label{corr_maps}
\end{figure*}

The results of stacking all (i.e., not selected by redshift) SCUBA,
24$\mu$m and 8$\mu$m sources onto the MIPS data are given in Table
\ref{tab_fluxes}. Offset maps showing the average weighted flux per
source for each combination of MIPS image and source list are also
plotted in Figure \ref{corr_maps}. Errors in both the table and the maps
include the uncertainty of the calibration on the 70 and 160$\mu$m
data.

\begin{table*}
\begin{tabular}{|l|c|c|c|}
\hline
MIPS Data & 850$\mu$m SCUBA Sources & 24$\mu$m Sources & 8$\mu$m Sources \\
\hline
70$\mu$m & 3.63$\pm$0.77 (3.45$\pm$0.80) & 2.04$\pm$0.25 (2.10$\pm$0.25) 
      & 0.95$\pm$0.12 (0.83$\pm$0.12) \\
70$\mu$m - 3$\sigma$ sources & 2.51$\pm$0.77 (2.58$\pm$0.78) 
    & 1.64$\pm$0.24 (1.65$\pm$0.24) & 0.64$\pm$0.12 (0.57$\pm$0.12) \\
160$\mu$m & 19.9$\pm$6.4 (16.9$\pm$6.6) &  15.0$\pm$2.4 (14.9$\pm$2.4) 
    & 8.5$\pm$1.1 (8.2$\pm$1.1) \\
160$\mu$m - 3$\sigma$ sources & 9.5$\pm$6.6 (8.1$\pm$6.7) 
    & 8.8$\pm$2.2 (8.6$\pm$2.2) &  5.6$\pm$1.0 (5.4$\pm$1.0) \\
\hline
\end{tabular}
\normalsize
\caption{Average weighted flux per stacked source ($\overline{f}_w$)
in mJy for the MIPS 70 and 160$\mu$m images, including and having
subtracted $\geq 3\sigma$ sources. These correspond to the flux at
$(0,0)$ in the offset maps shown in Figure \ref{corr_maps}. Quantities
in parentheses are the average unweighted fluxes $\overline{f}$. All
errors include the MIPS 70 and 160$\mu$m calibration uncertainty.}
\label{tab_fluxes}
\end{table*}

Apart from a single case, i.e., the instance in which SCUBA sources
were stacked onto the 160$\mu$m MIPS image with $\geq 3\sigma$ sources
removed, every stacking combination results in a significant detection
of far-IR flux. All peaks in the offset maps are well aligned with the
origin. This indicates that all data are properly aligned, as we
expected since all {\sl Spitzer} data are tied to 2MASS \citep[Two Micron
All Sky Survey; ][]{cutri03} and we showed in \citet{dye06} that the
SCUBA data are well aligned with the {\sl Spitzer} data. The more extended
nature of the peaks in the 160$\mu$m offset maps is a reflection of
the broader PSF at this wavelength (40\arcsec FWHM, compared to
18\arcsec at 70$\mu$m).

The differences between the $3\sigma$ source-subtracted and unmodified
stacks show that at both 70 and 160$\mu$m, the $\geq 3\sigma$ sources
account for approximately 50\% of the average flux per source, across
all three source populations.  In every case, the average flux per
SCUBA source is highest, followed by the average flux per 24$\mu$m
source then per 8$\mu$m source.  This is not surprising; the 70 and
160$\mu$m data are sensitive to the same dusty population of sources
as SCUBA, whereas the 8$\mu$m data are also sensitive to distant older
stellar populations. Also, the fact that there are more objects in the
24 and 8$\mu$m catalogues brings the average flux down because on
average, these sources will sample more image noise than areas of
significant 70 and 160$\mu$m emission.

\citet{dole06} stack 24$\mu$m sources with fluxes $\geq 60 \mu$Jy onto
MIPS data to measure a flux of $0.138\pm0.024$ MJy/sr at 70$\mu$m and
$0.571\pm0.123$ MJy/sr at 160$\mu$m. Using our MIPS data with all
$\geq 3 \sigma$ sources removed, we find a lower contribution of
0.070$\pm$0.010 MJy/sr at 70$\mu$m and 0.36$\pm$0.09 MJy/sr at
160$\mu$m. However, a contribution of 0.103$\pm$0.019 MJy/sr at
70$\mu$m and 0.87$\pm$0.16 MJy/sr at 160$\mu$m is made by the $8\mu$m
sources, again having removed the $\geq 3 \sigma$ sources. Within the
errors and including the fact that our data are more prone to cosmic
variance (see below) being $\sim 80$ times smaller in areal coverage,
our results are consistent with those of \citet{dole06}.

To estimate of the effects of cosmic variance on our results, we
divided the data into two approximately equal areas and then repeated
the stacking with the halved data. This was performed twice, firstly
splitting by the median source RA of each source catalogue, then by
the median Declination.  The $1\sigma$ variation in the spread of the
resulting stacked 70$\mu$m flux was found to be $\sim30\%$ and the
variation in the 160$\mu$m flux, $\sim20\%$.  Since this is an
estimate of the variance between fields half the size, the variance
between fields of the full size, discounting clustering effects, will
be a factor of $\sqrt{2}$ smaller, i.e., $\sim20\%$ at 70$\mu$m and
$\sim15\%$ at 160$\mu$m. As this serves merely as an
order-of-magnitude estimate of the cosmic variance, we do not include
it in any of the errors quoted in this paper.

\subsubsection{Contribution of sources to the CIB}

The total contribution of the three different source populations to
the CIB is given in Table \ref{tab_contrib}. By extrapolation,
\citet{chapman05} estimated an upper limit on the contribution of $>
1$mJy SCUBA sources detected at 850$\mu$m to the CIB emission at
160$\mu$m of $\simless 0.04$ MJy/sr.  Despite our CUDSS sources having
a brighter sensitivity level of 3.5mJy, we measure a higher
contribution at 160$\mu$m of 0.125$\pm$0.040 MJy/sr, without removing
any bright MIPS sources, or 0.060$\pm$0.042 MJy/sr having removed all
$\geq 3 \sigma$ sources. Although these measurements have large
uncertainties, they suggest the possibility of a somewhat larger SCUBA
source contribution to the 200$\mu$m CIB peak than previously thought.

Table \ref{tab_contrib} shows that at both 70 and 160$\mu$m, the SCUBA
sources make the lowest total contribution, followed by the 24$\mu$m
sources and then the 8$\mu$m sources with the highest
contribution. This is an important result; sources on the shorter
wavelength side of the peak in the CIB resolve more of the CIB at 70
and 160$\mu$m, and therefore most likely at the 200$\mu$m peak itself,
than the SCUBA sources on the longer wavelength side. This is almost
entirely due to the differing sensitivities of the source populations
used for stacking. In terms of the efficiency of resolving the bulk of
the CIB emission, the 8 and 24$\mu$m source population are more
favourable than the SCUBA sources. This is not surprising because
SCUBA surveys typically find 0.4 sources per arcmin$^2$ for each hour
of observation whereas {\sl Spitzer} surveys find $\sim130$ 8$\mu$m
sources per arcmin$^2$ for each hour of observation.  Of course, in
the context of the present study, SCUBA's time would be more
efficiently used by computing the cross correlation of the 850$\mu$m
maps with the {\sl Spitzer} images, rather than merely stacking at the
positions of significant SCUBA sources. We will carry out this cross
correlation in future work.

\begin{table}
\begin{tabular}{|l|c|c|c|}
\hline
MIPS Data & SCUBA & 24$\mu$m & 8$\mu$m \\
\hline
70$\mu$m   & 0.021$\pm$0.005 &  0.087$\pm$0.011 & 0.152$\pm$0.019  \\
70$\mu$m - 3$\sigma$ & 0.014$\pm$0.004 & 0.070$\pm$0.010 & 0.103$\pm$0.019  \\
160$\mu$m  & 0.125$\pm$0.040 &  0.62$\pm$0.10 & 1.31$\pm$0.17   \\
160$\mu$m - 3$\sigma$  & 0.060$\pm$0.042  & 0.36$\pm$0.09 & 0.87$\pm$0.16  \\
\hline
\end{tabular}
\normalsize
\caption{Contribution of sources to the CIB at 70 and 160$\mu$m in
units of MJy/sr, computed from the average weighted flux per source
given in Table \ref{tab_fluxes}. The coverage for each source
catalogue is as follows: 41 arcmin$^2$ for 850$\mu$m SCUBA sources, 49
arcmin$^2$ for the 24$\mu$m sources and 59 arcmin$^2$ for the 8$\mu$m
sources.}
\label{tab_contrib}
\end{table}

Expressing the absolute contributions in Table \ref{tab_contrib} as a
fraction of the CIB is somewhat difficult due to the uncertainty in
the background flux at 70 and 160$\mu$m (primarily because of
differing estimates of foreground contamination). In fact, there are
no direct measurements at these specific wavelengths.  The most
reliable measurements close to 160$\mu$m are those at 140$\mu$m made
by the Diffuse Infrared Background Experiment (DIRBE) on board the
Cosmic Background Explorer. Depending on the calibration used, the
DIRBE results give a flux of either $1.17 \pm 0.32$ MJy/sr or 0.70
MJy/sr at 140$\mu$m \citep{hauser98}, although as noted by
\citet{dole06}, the zodiacal cloud colours of \citet{kelsall98} imply
that a further 0.14 MJy/sr should be subtracted from these
numbers. Taking the average of both calibrations and subtracting 0.14
MJy/sr gives a flux of 0.80 MJy/sr. This can be extrapolated to give
an approximation of the flux at 160$\mu$m of 0.99 MJy/sr using the SED
fit to the CIB by \citet{fixsen98}. DIRBE also provided estimates of
the CIB at 60$\mu$m which are a useful constraint on our measurement
of the background at 70$\mu$m with MIPS. \citet{finkbeiner00} placed
an upper limit on the CIB at 60$\mu$m of $0.56 \pm 0.14$ MJy/sr. More
recently, this limit was reduced to 0.3 MJy/sr by \citet{dwek05}.

In terms of lower limits on the CIB at these wavelengths, stacking
analyses are currently the most stringent. By effectively
extrapolating their 24$\mu$m number counts, \citet{dole06} currently
provide the highest lower limits on the CIB of 0.17$\pm$0.03 MJy/sr at
70$\mu$m and 0.71$\pm$0.09 MJy/sr at 160$\mu$m.  Using these and
taking the DIRBE measurements quoted above as upper limits, we can
estimate the range in fractional contribution that our strongest
contributors, the 8$\mu$m sources, make to the CIB.  The upper limit
to the CIB at 60$\mu$m imposed by \citet{dwek05} and the extrapolated
lower limit of \citet{dole06} at 70$\mu$m indicates that our 8$\mu$m
sources resolve $\sim 35-75$\% of the background at 70$\mu$m.
Similarly, taking the DIRBE extrapolation to 160$\mu$m as an upper
limit and the extrapolated lower limit of \citet{dole06} at 160$\mu$m
implies that $\sim 90-100$\% of the 160$\mu$m background is resolved
by the 8$\mu$m sources.


To what extent do the additional 8$\mu$m sources not detected in the
24$\mu$m data emit at far-IR wavelengths? This can be very crudely
estimated by calculating the number of 8$\mu$m sources that would be
required to give the same measured CIB contribution but assuming each
source has a constant flux equal to the corresponding average 24$\mu$m
source flux. Here, the assumption is made that all the 24$\mu$m
sources (95\% of which are detected at 8$\mu$m) contribute to the
far-IR flux.  This simple calculation, shows that $\sim20\%$ and $\sim
50\%$ of the additional 8$\mu$m sources at 70 and 160$\mu$m
respectively would have to contribute in that case. This is a
conservative estimate because in reality, the average far-IR flux of
the additional 8$\mu$m sources will be lower than the average flux of
the 24$\mu$m sources due to the increased sensitivity of IRAC at
8$\mu$m.

\begin{figure}
\epsfxsize=8.3cm
{\hfill
\epsfbox{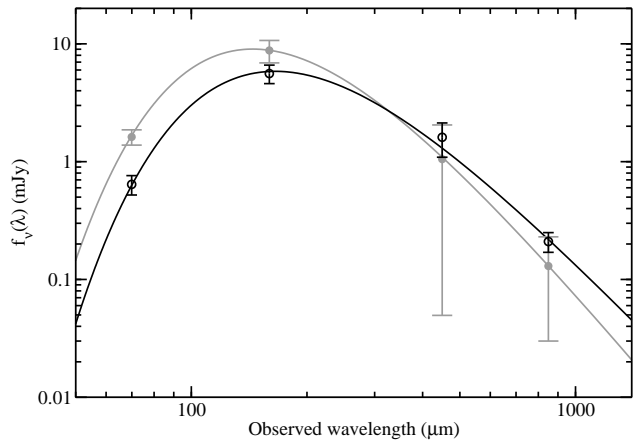}
\hfill}
\epsfverbosetrue
\caption{\small SED fits to the average 8$\mu$m (black line, open
circles) and 24$\mu$m (grey line, filled circles) source fluxes for
the median source redshift of $z=1$. Data points at 70 and 160$\mu$m
are taken from the present study and those at 450 and 850$\mu$m from
\citet{dye06}. Both SEDs show that the average 8 and 24$\mu$m sources
detected by {\sl Spitzer} in this dataset are borderline ULIRGs.}
\label{average_sed}
\end{figure}

\subsubsection{Average 8 \& 24$\mu$m source SEDs}
\label{sec_av_sed}

In \citet{dye06} we measured the average 450 and 850$\mu$m flux per 24
and 8$\mu$m source. Combining these measurements with the average 70
and 160$\mu$m flux per 24 and 8$\mu$m source determined in the present
work gives four data points each to which we can fit average
SEDs. Figure \ref{average_sed} shows the results of fitting the
grey-body function $A\nu^{\beta}{\rm B}(\nu,{\rm T})$ to these average
fluxes, where $A$ is a normalisation constant and B is the Planck
function.  In the fit, the parameters $A$, $\beta$ and T were allowed
to vary and we redshifted the function to the median redshift of our
sample, $z=1.0$.

For the 24$\mu$m sources, the best fit is achieved with
$\beta=2.05^{+1.03}_{-0.59}$ and T = $39.7^{+5.7}_{-5.4}$K and for the
8$\mu$m sources with $\beta=1.54^{+0.27}_{-0.28}$ and T =
$38.8^{+3.2}_{-2.7}$K ($1\sigma$ errors quoted).  The dependence of
these fitted parameters on the median redshift is such that a change
in redshift $\Delta z$ produces a change in T given by $\Delta {\rm T}
\simeq 19 \Delta z$ for both 24 and 8$\mu$m sources, while $\beta$ has
absolutely no dependence. The temperature of our average sources is
consistent with temperatures of submm galaxies found in the local
universe, e.g., \citet{dunne01} who measure T=$(36\pm5)$K. Whether
there is consistency with submm sources in the high redshift Universe
is less clear.  The sample of 73 SCUBA sources with a median redshift
of 2.3 of \citet{chapman05} has a median temperature of T$_{\rm med}
\simeq (36\pm7)$K, consistent with our values. However, the sample of
10 SCUBA sources with a median redshift of 1.7 of \citet{pope06} has a
lower median temperature of T$_{\rm med} \simeq 30$K.  If a
discrepancy exists, then it could be explained, at least in part, by
the selection effect noted by \citet{chapman05}; surveys like that of
\citet{pope06} requiring a submm and radio detection are biased toward
colder sources.

The normalisation of both SEDs confirms that the average 24 and
8$\mu$m source detected by {\sl Spitzer} in the current sample is a
borderline ultra-luminous infrared galaxy (ULIRG).  To demonstrate
this, we use the definition of \citet{clements99} that stipulates a
ULIRG must have a luminosity of at least $2.5\times10^{11}$L$_{\odot}$
measured at 60$\mu$m by the Infra-Red Astronomical Satellite (IRAS).
The rest-frame 60$\mu$m flux computed from our best fit SEDs is
$2.3\times10^{11}$L$_{\odot}$ for the average 24$\mu$m source and
$1.2\times10^{11}$L$_{\odot}$ for the average 8$\mu$m source, in
good agreement with \citet{dye06}.

An interesting question is how do our average 24 and 8$\mu$m sources
compare to the average SCUBA source detected by other studies? 
\citet{pope06} define the quantity L$_{\rm IR}$ as the integral of
flux in the wavelength range 8 - 1000$\mu$m.  Their sample of 10 SCUBA
sources has a median L$_{\rm IR}$ of $6.0\times
10^{12}$L$_{\odot}$. Similarly, the median value of L$_{\rm IR}$ for
the sample of 73 SCUBA sources of \citet{chapman05} is $8\times
10^{12}$L$_{\odot}$. In comparison, our average 24$\mu$m SED gives
L$_{\rm IR}=5.8\times 10^{11}$L$_{\odot}$ and our average 8$\mu$m SED
L$_{\rm IR}=3.5\times 10^{11}$L$_{\odot}$. The average 24 and 8$\mu$m
source in our sample is therefore $\sim 10$ times fainter than the
average SCUBA source detected by the previous two studies.

\subsection{Stacking the 24 and 8$\mu$m sources by redshift}

In this section, we consider the contribution from the 24 and 8$\mu$m
sources to the CIB at 70 and 160$\mu$m as a function of
redshift. Using the photometric redshifts already determined in
\citet{dye06}, we divided the sources equally into redshift bins of
varying width. The source redshifts extend up to $z\simeq 4$
\citep[see Figure 3 of][]{dye06}. Bins were chosen to be large
compared to the average redshift uncertainty but small enough to give
reasonable resolution, hence the 24$\mu$m sources were divided into 5
redshift bins and the 8$\mu$m sources into 6. Approximately 10\% of
sources with undetermined redshifts (due to their sparse optical
photometry) were omitted from the analysis of this section. This
therefore gives $\sim 30$ objects per 24$\mu$m bin and $\sim 120$
objects per $8\mu$m bin.

\begin{figure}
\epsfxsize=8cm
{\hfill
\epsfbox{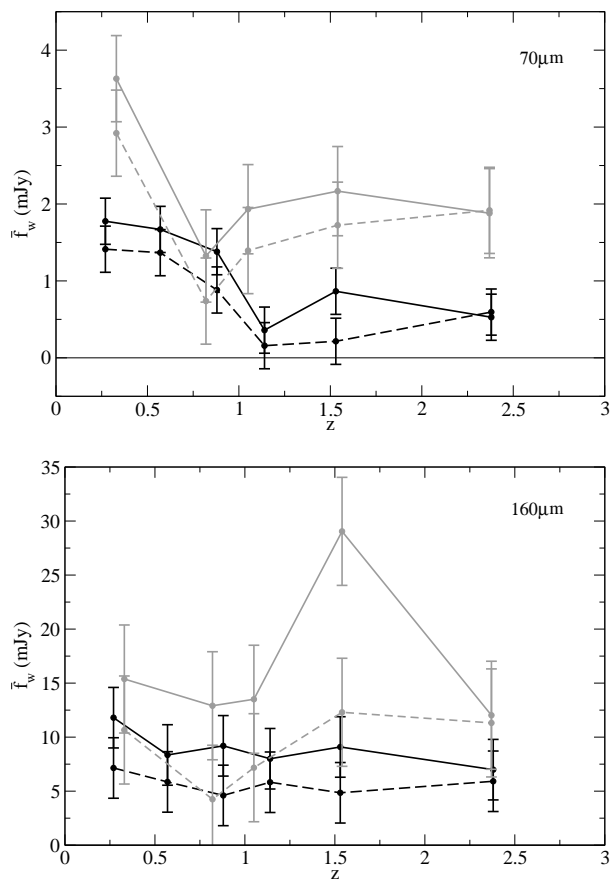}
\hfill}
\caption{Variation of average weighted 70$\mu$m flux (top) and
160$\mu$m flux (bottom) of {\sl Spitzer} 24 and 8$\mu$m objects binned by
redshift. The median redshift in each bin is plotted. Redshifts extend
up to $z\simeq 4$ \citep[see][]{dye06} and are divided equally between
bins. In both plots, the continuous grey and black lines correspond to
the 24 and 8$\mu$m sources respectively, stacked onto the MIPS images
without any sources removed. The dashed lines show the average
weighted flux having removed the $\geq 3\sigma$ sources from the MIPS
images.}
\label{f_vs_z}
\end{figure}

Figure \ref{f_vs_z} shows how the weighted average 70 and 160$\mu$m
flux per source varies with redshift. The plots show this variation
both having removed the $\geq 3\sigma$ sources in the MIPS images and
with them left in place.  At 70$\mu$m, the effect of removing the
3$\sigma$ sources has less effect than at 160$\mu$m.  Also, at
70$\mu$m, the flux is dominated by 8$\mu$m sources lying at lower
redshifts. Dividing the 8$\mu$m sources into two populations
segregated by the median redshift, $z=1.0$, the low redshift
population accounts for $(79\pm10)\%$ of the total 70$\mu$m emission
(having removed the $\geq 3 \sigma$ sources) from the 8$\mu$m
sources. In comparison, the 70$\mu$m emission per 24$\mu$m source is
more evenly spread in redshift, the low redshift population accounting
for $(51\pm10)\%$.  At 160$\mu$m, the low redshift 8$\mu$m sources
contribute $(52\pm14)\%$ of their total and the 24$\mu$m sources
contribute $(42\pm15)\%$, having removed all $\geq 3 \sigma$ 160$\mu$m
sources.

The differences between the 70 and 160$\mu$m plots in Figure
\ref{f_vs_z} are very well explained by a single average source SED
consistent with the fitted average SEDs derived in section
\ref{sec_av_sed}. To demonstrate this, we took an SED from
\citet{dale02} corresponding to a dust temperature of T=40K to match
our average SEDs. Since this SED extends into the optical and models
typical mid-IR spectral features due to dust, a realistic prediction
of the 70 and 160$\mu$m flux of a source given its redshift and 8 or
24$\mu$m flux can be made. In this way, using our 8 and 24$\mu$m source
catalogues, we computed a prediction of the variation of 70 and
160$\mu$m flux with redshift.

\begin{figure}
\epsfxsize=8cm
{\hfill
\epsfbox{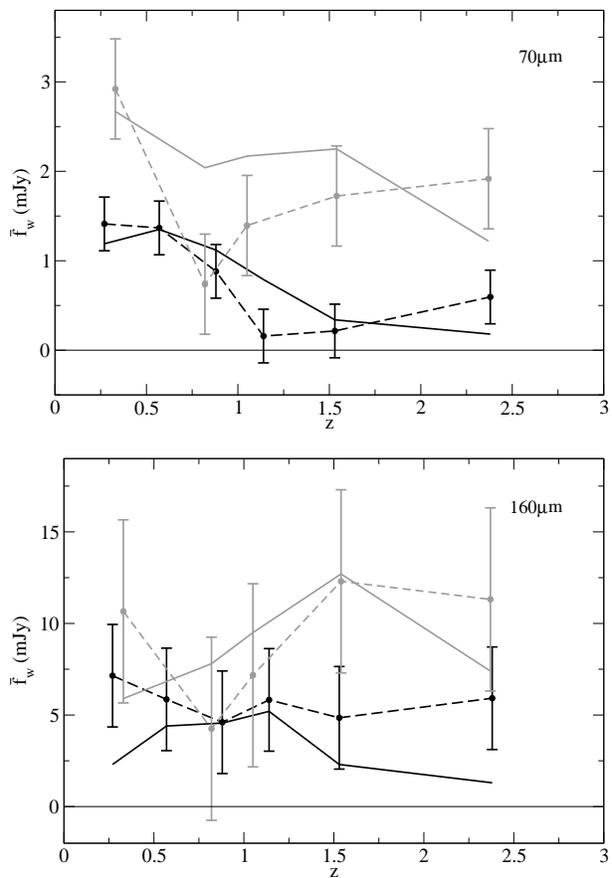}
\hfill}
\caption{Comparison of the measured (dashed lines) and predicted
(continuous lines) MIPS 70 and 160$\mu$m flux from the 8$\mu$m (black) and
24$\mu$m (grey) sources using a T=40K SED taken from
\citet{dale02}. The measured data are the stacked fluxes in Figure
\ref{f_vs_z} with $\geq 3\sigma$ sources removed.}
\label{f_vs_z_predicted}
\end{figure}

The results of this analysis are shown in Figure
\ref{f_vs_z_predicted}.  There is good agreement between our measured
variation and the predicted variation. We reproduce the total flux
(summed over all sources) within the errors and also the observed
trends. Most notably, we reproduce the decline in the 70$\mu$m
emission from the 8$\mu$m sources within $0<z<1$ as the peak of the
SED is redshifted out of the 70$\mu$m band. This explains why the
majority of 70$\mu$m emission is observed from the 8$\mu$m sources
lying at $z\leq1$.  The prediction degrades quickly if a cooler or
warmer SED is used; with a 35K or 45K SED, the total predicted flux is
inconsistent with the total measured. The fact that a single SED can
be used to fit the observed flux over such a wide range of redshifts
implies that only a small amount of source evolution must have occured
during that time.

To assess the effects of cosmic variance on the results of this
section, we repeated the previous exercise of dividing the data into
halves and re-stacking.  We found that the major trends are robust,
i.e., the decline in 70$\mu$m flux from the 8$\mu$m sources over the
interval $0<z<1$ and that the other combinations remain consistent
with little or no variation with redshift. However, the large spike in
160$\mu$m flux seen from the 24$\mu$m sources at $z\sim 1.5$ (without
having removed the $\geq 3 \sigma$ sources) is not robust and
therefore presumably an effect of cosmic variance.

\section{Summary and Discussion}
\label{sec_summary}

In this paper, we have quantified the contribution of flux to the CIB
at 70 and 160$\mu$m from SCUBA sources and 24 and 8$\mu$m {\sl Spitzer}
sources in the CUDSS 14hour field. By stacking flux at the position of
the different sources, we have found that the SCUBA sources make the
highest contribution per source and that the 8$\mu$m sources make the
lowest. Conversely, the opposite is true of the total contribution
from all sources, leading to the conclusion that the bulk of the CIB
is most efficiently resolved by sources detected at wavelengths
shorter than the peak of the CIB emission at $\sim 200\mu$m.

Our stacking suggests a somewhat larger contribution from the CUDSS
SCUBA sources to the 200$\mu$m CIB peak than previously
thought. \citet{chapman05} estimated an upper limit on the
contribution of $> 1$mJy SCUBA sources detected at 850$\mu$m to the
CIB emission at 160$\mu$m of $\simless 0.04$ MJy/sr.  Despite our
SCUBA sources having a brighter sensitivity level of 3.5mJy, we
measure a contribution at 160$\mu$m of 0.125$\pm$0.040 MJy/sr, without
removing any bright MIPS sources, or 0.060$\pm$0.042 MJy/sr having
removed all $\geq 3 \sigma$ sources. 

Since measurements of the CIB at 70 and 160$\mu$m are presently very
uncertain, the fractional contribution to the CIB made by our sources
can only be expressed within a range set by present upper and lower
limits. Using the DIRBE estimates as upper limits and the lower limits
set by \citet{dole06}, our strongest contributors, the 8$\mu$m
sources, resolve somewhere between $\sim 35-75$\% of the background at
70$\mu$m and $\sim 90-100$\% at 160$\mu$m.

By combining our results in the present work with our previous
stacking of 450 and 850$\mu$m SCUBA flux \citep{dye06}, we have
established that the 8 and 24$\mu$m sources detected by {\sl Spitzer}
are on average borderline ULIRGs. The average source we detect is
$\sim 10$ times fainter than the average SCUBA source detected by
\citet{pope06} and \citet{chapman05} integrating flux over the
wavelength range 8 - 1000$\mu$m. Furthermore, the temperature of $\sim
40$K of our average source is consistent with temperatures of submm
galaxies found in the local Universe \citep[e.g.,][]{dunne01} and with
the median temperature of $(36\pm7)$K of SCUBA sources in the high
redshift Universe found by \citet{chapman05}. However, our average
source is warmer than the median temperature of T$_{\rm
med} \simeq 30$K of 10 SCUBA sources measured by \citet{pope06}.

Using photometric redshifts assigned to the 8 and 24$\mu$m sources, we
have investigated how the contribution of 70 and 160$\mu$m flux to the
CIB varies with redshift. We have found that the 8$\mu$m sources at
low redshifts, $z < 1$ (accounting for half of them), are the
strongest contributors to the CIB at 70$\mu$m, their flux amounting to
$\sim 4$ times that of the $z>1$ sources.  The 70$\mu$m emission per
24$\mu$m source as well as the 160$\mu$m emission per 8 and 24$\mu$m
source is consistent with an even distribution over redshift. This
verifies the result of \citet{dole06} that the majority of
the emission at 70 and 160$\mu$m from 24$\mu$m sources must come from
a redshift of $z \sim 1$ where the redshift distribution of these
sources peaks. We have shown how this distribution can be reproduced
from our observed 8 and 24$\mu$m catalogue of fluxes and redshifts
using a single non-evolving source SED with a dust temperature of 40K.

As a concluding remark, this study and similar recent studies
\citep[e.g.,][]{serjeant04,knudsen05,dye06,wang06} indicate that the
CIB is not predominantly due to a rare population of exceptionally
luminous submillimeter sources as hinted at by early SCUBA
observations, but that a much more numerous galaxy population of
modest average luminosity is responsible instead. However, we have
shown here that the SCUBA galaxies probably do make a larger
contribution than previously thought, although much larger numbers of
SCUBA sources such as those of the SCUBA half degree extragalactic
survey \citep[SHADES; ][]{mortier05} or those resulting from future
SCUBA2 surveys\footnote{see http://www.roe.ac.uk/ukatc/projects/scubatwo} will
be required to improve the precision of these measurements.

\appendix
\section{Flux boosting correction}
\label{app_flux_deboost}

In the following, it is assumed that the MIPS image has been prepared
such that the value of any one pixel gives the total flux a point
source would have if located within that pixel. Since this preparation
inevitably involves convolution of the raw MIPS image with some kind
of kernel, the profile of a point source will be the convolution of
this kernel with the original image PSF. This resulting profile is
referred to as the image `beam' hereafter.

Suppose that $T$ is the total number of sources being stacked onto the
MIPS image and that a subset $N$ of these are `genuine' sources. A source
is defined as `genuine' if it has associated MIPS emission that makes a
non-negligible contribution to the stacked flux. (In practice, one
would derive a threshold flux that depends on the number of
sources). The actual average flux per source is then simply the summed
flux from the genuine sources divided by the total number of sources,
\be
\label{eq_f_actual}
\overline{f}_{\rm actual} = \frac{1}{T}\sum_{i}^{N} f_i \,\, ,
\ee
where $f_i$ is the flux in the MIPS image pixel at the position of the
genuine source $i$.

However, this quantity is overestimated if one naively adds the flux in
the MIPS image at the positions of all $T$ sources for two reasons.
Firstly, the $T-N$ sources without associated MIPS emission (the
`contaminating' sources) sample flux from the genuine sources since some
will happen to lie within genuine source beams. The extra flux sampled
on average from a genuine source with flux $f_i$ by the contaminating 
sources is
\ba
f^{c}_{i} =  2\pi n_c f_i \int r b(r) {\rm d}r = n_c B_e f_i,
\ea
where $b(r)$ is the radial beam profile scaled to have
a peak height of unity, $n_c$ is the number density
of contaminating sources and $B_e$ defines the effective beam
area. Secondly, the genuine sources themselves sample emission from
neighbouring genuine sources when their beams overlap. Similar to the
contaminating sources, the extra flux sampled on average from a genuine
source with flux $f_i$ by its neighbouring genuine sources is
\be
f^{g}_{i}=n'_g B_e f_i  \,\, ,
\ee 
where $n'_g$ is the number density of the neighbouring
$N-1$ genuine sources.
The average MIPS flux per source that is measured by
summing the flux at all $T$ source positions is therefore
\ba
\label{eq_f_measured}
\overline{f}_{\rm measured} 
&=& \frac{1}{T} \sum_{i}^{N} (f_i + f^{c}_{i} + f^{g}_{i}) 
= \frac{1+ n'_t B_e}{T} \sum_{i}^{N}f_i \nonumber \\
           &=& (1+n'_t B_e) \overline{f}_{\rm actual}\,\, ,
\ea
where $n'_t=n'_g+n_c$ is the number density of $T-1$ sources.
The measured average flux is therefore the actual
average flux boosted by a factor of $(1+n'_t B_e)$.

If the MIPS data are properly normalised, (so that there is
no net positive emission from artifacts, systematic effects, etc.)
then the average of the image over its area $A$ is
\be
\label{eq_f_image}
\overline{f}_{\rm image} = \frac{B_e}{A} \sum_{i}^{N} f_i
= \frac{n_t B_e}{T} \sum_{i}^{N} f_i
= n_t B_e \overline{f}_{\rm actual}  \,\, ,
\ee 
having used the fact that the number density of {\em all} $T$
sources is $n_t=T/A$. Subtracting equation (\ref{eq_f_image}) from
equation (\ref{eq_f_measured}) and rearranging gives
\be
\overline{f}_{\rm actual}=(\overline{f}_{\rm measured} -
\overline{f}_{\rm image})(1-B_e/A)^{-1} \,\, ,
\ee
hence the actual average flux per source can be obtained by
subtracting the image average and dividing by the factor $(1-B_e/A)$.

In the above, it has been assumed that source positions are
a random sampling of a uniform distribution. In reality, the
sources will exhibit a degree of clustering. Clustering of
the genuine sources causes a positive bias of the average flux
compared to the non-clustered assumption, whereas clustering of
the contaminating sources on average has no effect.

\begin{flushleft}
{\bf Acknowledgements}
\end{flushleft}

SD is supported by the U.K. Particle Physics and Astronomy Research
Council. Part of the research described in this paper was carried out
at the Jet Propulsion Laboratory, California Institute of Technology,
under a contract with the National Aeronautics and Space
Administration.  The Spitzer Space Telescope is operated by the Jet
Propulsion Laboratory, California Institute of Technology under NASA
contract 1407. Support for this work was provided by NASA through
contract 1256790 issued by JPL/Caltech. We thank Herve Dole for
refereeing this work and providing several helpful suggestions which
have enhanced this paper.


{}

\label{lastpage}

\end{document}